\documentclass[aps,prl,twocolumn,amssymb]{revtex4}

\usepackage{graphicx}

\begin{document}

\pagestyle{empty}
{\bf \noindent Comment on ``Probing the equilibrium dynamics of
colloidal hard spheres above the mode-coupling glass transition''}

In the Letter \cite{Bram09}, Brambilla {\it et al.} claimed to observe activated dynamics in  colloidal hard spheres above the critical packing fraction $\varphi_c$ of mode coupling theory (MCT). By performing microscopic MCT calculations, we show that 
polydispersity in their system shifts $\varphi_c$ above the
value determined by van Megen {\it et al.} for less polydisperse samples \cite{Meg98}, and that the data agree with theory except for, possibly, the highest $\varphi$. 

Brambilla {\it et al.} performed dynamic light scattering (DLS) on particles with average diameter $\bar d=210 $nm, and size polydispersity  $\sigma_{\rm pol}\approx 12.2$\% \cite{Meg10,Bram10}. They purport to measure $g_1=w\, \Phi^s_q(t)+ (1\!\!-\!\!w)\, \Phi_q(t)$ at wavevector $q=5.25/\bar d$, where  $\Phi_q(t)$ ($\Phi^s_q(t)$) are the coherent (incoherent) intermediate  scattering functions, and $w\approx0.8$.

 
It is generally expected, that the relaxation time $\tau$ increases with $\varphi$. However, several pairs of measured $\tau$  violate this requirement; see inset (a). Based on this, we estimate an error  in $\varphi$ of about $\pm0.002$.

Using MCT, we 
calculated the (in-) coherent correlators  of (for simplicity)  monodisperse  hard spheres at $qd=5.25$ in Percus-Yevick  approximation following
Ref.~\cite{Voig03}. The mixing parameter $w=0.53$ and the short time diffusivity  $D_s=0.59 \mu m^2/$s were fixed by matching the amplitude of the final and the rate of the initial decay, respectively. The fits in Fig.~1 result from varying the packing fraction $\varphi^{\rm mct}$  as shown in inset (b). The correlators close to $\varphi_c$  are described well \cite{note}, polydispersity shifts the glass transition upwards by 4\% to $\varphi_c=0.595$ \cite{Meg98}, and
the relation between experimental packing fraction $\varphi$ and its
fitted value in MCT is linear within the afore estimated error. The linearization predicts a power-law divergence of the relaxation time, $\tau\propto (\varphi_c-\varphi)^{-2.46}$, which is compatible with the measured relaxation times; see inset (c).

Brambilla {\it et al.}
interpret the scatter of the data around the theoretical lines above  $\varphi=0.585$ as indication of activated processes not contained in MCT. We find that the MCT fit only fails to explain the final decay at the highest density. We base this on the uncertainty in the packing fraction, $\Delta \varphi=0.002$, estimated beforehand.


\begin{figure*}
\includegraphics[width=0.9\linewidth]{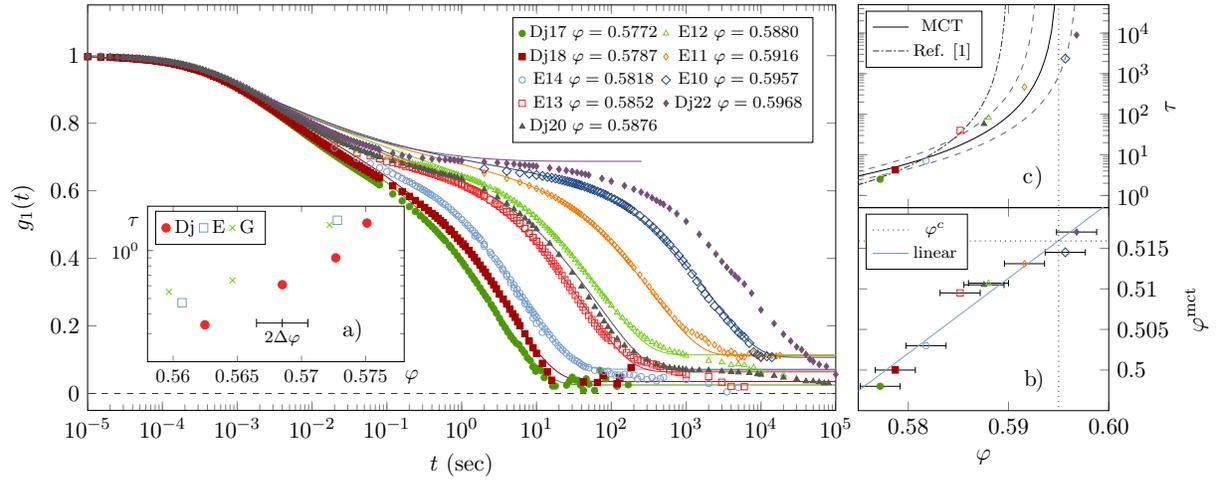}
\caption{Correlators from DLS \cite{Bram09} at packing fractions $\varphi$ in two dilution series (abbr. Dj and E) and from MCT for fitted $\varphi^{\rm mct}$ given in inset (b). The fits lead to $\varphi_c=0.595$, indicating that only Dj22 relaxes by activation not captured in MCT. The expected linear relation $\varphi^{\rm mct}-\varphi$ holds within errors, which  are estimated as $\pm\Delta \varphi= \pm0.002$ from the non-montonous ordering of $\tau - \varphi$ data (see inset a). The linear density relation gives a power-law divergence of  $\tau$ at $\varphi_c$ (solid line in inset c), respectively, at $\varphi_c\pm\Delta\varphi$ (dashed lines)  accounting for the density uncertainty; the dot-dashed curve is reproduced from \cite{Bram09}. }
\end{figure*}

\smallskip

\noindent
J. Reinhardt, F. Weysser, and M. Fuchs\\
\indent {\small
Fachbereich Physik, Universit\"at Konstanz, \\\indent
 78457  Konstanz, Germany\\}
\indent {\small
e-mail: matthias.fuchs@uni-konstanz.de}
\smallskip

\noindent
Received {\today}\\
\noindent
DOI:


\begin{thebibliography}{10}
\bibitem{note} Base-lines were estimated as shown. Correlators in \cite{Bram09}, whose plateau height was 
not matched to short-time data and where crossings exist due to experimental uncertainties in the short-time matching, 
were neglected.
\end{thebibliography}
\end{document}